\documentclass[prb,superscriptaddress,preprint, floats]{revtex4}

\usepackage{amsmath, amsthm, amssymb, pifont, wasysym}

\usepackage{graphicx}
\usepackage{bm} 
\usepackage{pifont}
\usepackage{dcolumn} 
\usepackage{color}

\newcommand {\Jex}{$J_{\mathrm{ex}}$}

\newcommand {\syxA}{$\sigma_{\mathrm{yx, A}}$}
\newcommand {\sxxint}{$\sigma_{\mathrm{xx, int}}$}
\newcommand {\syxAint}{$\sigma_{\mathrm{yx, A, int}}$}
\newcommand {\RyxA}{$R_{\mathrm{yx, A}}$}
\newcommand {\Oz}{$\Omega^{\mathrm{z}}_{n,\bm{k}}$}
\newcommand {\sz}{$\sigma_{\mathrm{z}}$}
\newcommand {\EF}{$E_{\mathrm{F}}$}
\newcommand {\ED}{$E_{\mathrm{D}}$}

\newcommand {\CA}{Cd$_3$As$_2$}

\newcommand {\NB}{Na$_3$Bi}

\newcommand {\TiO}{TiO$_2$}
\newcommand {\SN}{Si$_3$N$_4$}

\newcommand {\BTIG}{(Bi,Tb)$_3$Fe$_5$O$_{12}$}
\begin{document}

\title{Ferromagnetic state above room temperature \\ in a proximitized topological Dirac semimetal}

\author{Masaki Uchida}
\email[Author to whom correspondence should be addressed: ]{uchida@ap.t.u-tokyo.ac.jp}
\affiliation{Department of Applied Physics and Quantum-Phase Electronics Center (QPEC), University of Tokyo, Tokyo 113-8656, Japan}
\affiliation{PRESTO, Japan Science and Technology Agency (JST), Tokyo 102-0076, Japan}
\author{Takashi Koretsune}
\affiliation{Department of Physics, Tohoku University, Sendai 980-8578, Japan}
\author{Shin Sato}
\affiliation{Department of Applied Physics and Quantum-Phase Electronics Center (QPEC), University of Tokyo, Tokyo 113-8656, Japan}
\author{Markus Kriener}
\affiliation{RIKEN Center for Emergent Matter Science (CEMS), Wako 351-0198, Japan}
\author{Yusuke Nakazawa}
\affiliation{Department of Applied Physics and Quantum-Phase Electronics Center (QPEC), University of Tokyo, Tokyo 113-8656, Japan}
\author{Shinichi Nishihaya}
\affiliation{Department of Applied Physics and Quantum-Phase Electronics Center (QPEC), University of Tokyo, Tokyo 113-8656, Japan}
\author{Yasujiro Taguchi}
\affiliation{RIKEN Center for Emergent Matter Science (CEMS), Wako 351-0198, Japan}
\author{Ryotaro Arita}
\affiliation{Department of Applied Physics and Quantum-Phase Electronics Center (QPEC), University of Tokyo, Tokyo 113-8656, Japan}
\affiliation{RIKEN Center for Emergent Matter Science (CEMS), Wako 351-0198, Japan}
\author{Masashi Kawasaki}
\affiliation{Department of Applied Physics and Quantum-Phase Electronics Center (QPEC), University of Tokyo, Tokyo 113-8656, Japan}
\affiliation{RIKEN Center for Emergent Matter Science (CEMS), Wako 351-0198, Japan}

\begin{abstract}

We report an above-room-temperature ferromagnetic state realized in a proximitized Dirac semimetal, which is fabricated by growing typical Dirac semimetal {\CA} films on a ferromagnetic garnet with strong perpendicular magnetization. Observed anomalous Hall conductivity with substantially large Hall angles is found to be almost proportional to magnetization and opposite in sign to it. Theoretical calculations based on first-principles electronic structure also demonstrate that the Fermi-level dependent anomalous Hall conductivity reflects the Berry curvature originating in the split Weyl nodes. The present Dirac-semimetal/ferromagnetic-insulator heterostructure will provide a novel platform for exploring Weyl-node transport phenomena and spintronic functions lately proposed for topological semimetals.

\end{abstract}
\maketitle

Topological materials, marked by non-trivial topology of electronic band dispersions, have demonstrated great potential for investigating relativistic electron dynamics in solids. Due to the bulk band inversion, topological insulators host surface states with Dirac dispersion and helical spin texture \cite{TIreview1}. As another leading example, topological semimetals have received increasing attention in recent years \cite{TDSreview1, TDSreview2, topotransition, theoryTDS, theoryCA, theoryNB, ARPESCA1, ARPESCA2, weylTaAs}. In topological Dirac semimetals, their conduction and valence bands touch at point nodes protected by crystalline symmetry and form bulk three-dimensional Dirac cones \cite{topotransition, theoryTDS, theoryCA, theoryNB, ARPESCA1, ARPESCA2}. The Dirac node with double degeneracy is a superposition of Weyl nodes with opposite chirality, which separately appear in Weyl semimetals \cite{topotransition, weylTaAs}. The Weyl nodes with plus/minus chirality correspond to source/drain of the Berry curvature field and act as synthetic plus/minus magnetic monopoles in the momentum space \cite{SrRuO3}.

Topological electronic structures coupled with magnetization can produce various emergent magnetotransport originating in the Berry curvature. Intrinsic anomalous Hall effect is one typical example. A semiclassical picture of the anomalous Hall effect has been reexamined in the Berry-phase formulation \cite{intrinsicAHE1, AHE}, where the intrinsic anomalous Hall conductivity is expressed as Brillouin zone integral of the Berry curvature weighted by the band occupation. Namely, the momentum-space Berry-phase contribution is directly obtained from the semiclassical spin-orbit Hamiltonian introduced by Karplus and Luttinger \cite{KL}. Its quantization in magnetic topological insulators can be also considered as the quantization of the integral of the Berry curvature in two dimensions \cite{TIdopingQAHE2}. In the minimal model of magnetic Weyl semimetals, it has been theoretically shown that the anomalous Hall conductivity is saturated to a semiquantized value proportional to the separation of the Weyl nodes as the Fermi level approaches the node energy \cite{AHEinmagneticWeyl1, AHEinmagneticWeyl2}. Moreover, unique spintronic functions, such as electric-field-induced domain wall motion \cite{magneticWeylApp1} and spin torque generation \cite{magneticWeylApp3}, have been theoretically proposed for these simple magnetic Weyl semimetals, where the intrinsic anomalous Hall conductivity or the separation of the Weyl nodes simply emerges as a coefficient of the nontrivial coupling between the charge and the magnetization.

Actually, bulk magnetic Weyl semimetals have been intensively studied, producing a major breakthrough in the solid state physics \cite{Co3Sn2S2,Co3Sn2S2_suggested,Fe3GeTe2,CeAlGe1,CeAlGe2,PrAlGe_suggested,SrRuO3, Mn3Sn,GdPtBi, GdPtBi_chiralanomaly,RPtBi_suggested,Heusler_design}. On the other hand, most of their band structures are rather complicated. In particular, other bands crossing the Fermi level or multiple Weyl points dispersed in the momentum space should give complicated contribution to electronic transport, making it difficult to clearly demonstrate the Weyl-node based transport phenomena and spintronic functions. The ferromagnetic transition temperatures have been also limited to far below room temperature. In this light, systems with more simple and ideal band structures as well as higher ordering temperature have been strongly desired.

The Weyl nodes can be created also by inducing magnetization in topological Dirac semimetals. The Dirac semimetals are a highly suitable system for investigating Weyl-node based magnetotransport, because the symmetrically protected Dirac node can be continuously separated and controlled as a pair of the Weyl nodes in the momentum space, in proportion to the induced magnetization. As exemplified in typical Dirac semimetal materials {\CA} and {\NB}, it is also highly advantageous that their band structures composed of strongly dispersive $s$ or $p$ bands host only the simple Dirac dispersions over a wide energy range around the Fermi level \cite{theoryCA, ARPESCA1, ARPESCA2, theoryNB, QPI, highmobility, sigmaxx_suggested}, enabling us to more directly evaluate and harness potential of the Weyl nodes. In general, on the other hand, the high Fermi velocity of the Dirac dispersion makes it rather difficult to induce ferromagnetism to the bands. It has been highly challenging to realize the ferromagnetic phase from topological Dirac semimetal materials, even by magnetic doping such as of 3$d$ transition metal elements \cite{Crdoping1, Mndoping1}.

Here we report a robust ferromagnetic state realized by heterostructuring topological Dirac semimetal. High crystallinity {\CA} thin films with controlled carrier densities are grown on a rare-earth iron garnet substrate with large perpendicular magnetic anisotropy, as shown in Figs. 1(a) and (b). By inducing exchange splitting to the Dirac dispersion through magnetic proximity effect as utilized for topological insulators \cite{EuSBS3, BCSYIG, BSYIG, CBSTCS, BSTYIG1, BSTTIG, review}, an above-room-temperature ferromagnetic state is stabilized at the heterointerface. As the Weyl-node contribution to the anomalous Hall conductivity is demonstrated by theoretical calculations, this simple Dirac semimetal proximitized with the garnet will be a promising system for further studying Weyl-node based spintronic functions.

{\CA} films were epitaxially grown on {\BTIG} (111) single crystalline substrates by combining pulsed laser deposition and solid phase epitaxy techniques  \cite{PLDfilm1, PLDfilm2}. The film thickness is typically designed to be 60 nm to maintain the three-dimensional Dirac semimetal phase with avoiding the quantum confinement \cite{PLDfilm1, weylorbit2, weylorbit4}. The Bi substitution amount in the substrate was about 50\%. A Cd$_3$As$_2$ polycrystalline target was made by mixing 6N5 Cd and 7N5 As shots at a stoichiometric ratio of 3:2, reacting the mixture at 950 $^{\circ}$C for 48 hours in a vacuum-sealed tube, and then sintering the compound at 250 $^{\circ}$C for 30 hours after pelletization. The {\CA} films were deposited through a stencil metal mask at room temperature and patterned into a Hall bar shape with a channel width of 60 $\mu$m, as shown in the inset of Fig. 1(a). After the deposition, they were capped by 10 nm {\TiO} and 150 nm {\SN} in the chamber and then annealed at 600 $^{\circ}$C for 5 minutes in air. (112)-oriented {\CA} films were formed through high-temperature crystallization. Carrier density of the films was systematically controlled for samples A, B, and C ($n=1.6$, $1.0$, and $0.6\times10^{18}$ cm$^{-3}$). Longitudinal resistance $R_{\mathrm{xx}}$ and Hall resistance $R_{\mathrm{yx}}$ were measured with a standard four-probe method feeding a DC current of 3 $\mu$A on the multi-terminal Hall bar. Aluminum wires were connected to the Hall bar edges by breaking the capping layers using a wire bonding machine, and then reinforced by silver paste. The magnetotransport measurements were performed between 2 and 400 K in a Quantum Design Physical Properties Measurement System cryostat equipped with a 9 T superconducting magnet. The anomalous Hall resistance {\RyxA} was extracted by subtracting a linear ordinary component above 0.5 T. Magnetization $M$ and its temperature dependence were taken with applying a magnetic field perpendicular to the substrate plane. The measurements were performed by a superconducting quantum interference device magnetometer in a Quantum Design Magnetic Property Measurement System. $M$ parallel to the substrate plane was also measured for comparison \cite{supplemental}. Anomalous Hall conductivity {\syxA} was calculated based on first-principles electronic structure of $I4_{1}cd$ {\CA} \cite{theoryCA}. VASP and Wannier90 codes were used for extracting Cd $s$, $p$ and As $p$ Wannier orbitals. {\Jex}{\sz} is added for all these orbitals and {\syxA} was calculated using the Berry curvature formula (Eq. 1) at $k$-point grid $60\times60\times60$ and $T=100$ K. Here the $z$-axis is taken along the [112] direction. 

Figure 1(c) shows magnetoresistance measured for the Dirac-semimetal/ferromagnetic-insulator heterostructure. The main results are obtained for sample B (see also Supplemental Material for detailed transport properties of all the samples \cite{supplemental}). In addition to Shubnikov-de Haas oscillations resulting from the large Fermi velocity, a butterfly shaped hysteresis loop is observed around zero field, suggesting that magnetization is induced at the heterointerface and aligned along the [112] perpendicular direction above about 0.4 T. In this ferromagnetic state, the three-dimensional Dirac dispersion is expected to be altered as illustrated in Figs. 1(d)--(g). Each Dirac node separates into a pair of the Weyl nodes with opposite chirality (W$^{+}$ and W$^{-}$) along the $k_{\mathrm{[112]}}$ direction. While the induced magnetization may be dependent on the depth, such a position dependence is treated as a perturbation with keeping the Weyl node picture \cite{xdepreview}, resulting in a change in the distance ($2k_{\mathrm{W}}$) along the depth direction. Finite $k_{\mathrm{[112]}}$ ranges, where the Chern number of the two-dimensional slice is nonzero ($C=+1$), appear between the split Weyl nodes, accompanied by the breaking of time reversal symmetry. The Weyl nodes, once formed, are expected to stably exist until the pair annihilation.

Figure 2 summarizes anomalous Hall resistance {\RyxA} observed for the heterostructure at various temperatures. The garnet substrate, originally designed as a faraday rotator, shows sharp perpendicular magnetization and also hysteresis loop especially at low temperatures, while magnetic domains with reverse magnetization are formed before inverting the field direction \cite{supplemental}. In line with the perpendicular magnetization of the substrate, a clear anomalous Hall effect is detected with magnetic hysteresis. The saturation in {\RyxA} once vanishes at the magnetic compensation temperature of the substrate (130 K), but recovers and remains up to 400 K without changing its sign relative to the magnetization $M$. The exchange splitting is strongly induced by the magnetic proximity effect at the high quality heterointerface, realizing the robust ferromagnetic state in the {\CA} film.

The interface system with longitudinal conductivity $\sigma_{\mathrm{xx,int}}\sim 90$ $\Omega^{-1}\mathrm{cm}^{-1}$ is located in the dirty regime, where the anomalous Hall conductivity {\syxAint} is intrinsically induced by the anomalous velocity originating from the Berry phase, and also affected by impurity scattering with $\sigma_{\mathrm{xx,int}}$ \cite{intrinsicAHE1}. Assuming a proximitized thickness of 2 nm \cite{EuSBS3}, it is calculated to be $\sigma_{\mathrm{yx,A,int}}\sim10$ $\Omega^{-1}\mathrm{cm}^{-1}$. This assumption is reasonable to roughly estimate the interface contribution (for details see also Supplemental Material \cite{supplemental}). In the simplest case, where only a pair of the Weyl nodes exists at the Fermi level, the intrinsic anomalous Hall conductivity can be semiquantized as expressed in the following formula \cite{AHEinmagneticWeyl1, AHEinmagneticWeyl2}.
\begin{eqnarray}
\sigma_{\mathrm{yx, A}}=\frac{e^2}{\hbar}\int_{\mathrm{BZ}}\frac{d^{3}\bm{k}}{(2\pi)^3}\sum_{n}f(E_{n,\bm{k}})\Omega^{\mathrm{z}}_{n,\bm{k}}=-\frac{k_\mathrm{W}}{\pi}\frac{e^2}{h}
\end{eqnarray}
Here $f(E_{n,\bm{k}})$ is the Fermi-Dirac distribution for the $n$-th energy eigenvalue, $\Omega^{\mathrm{z}}_{n,\bm{k}}$ is $z$ component of the Berry curvature, and $k_{\mathrm{W}}$ is a half of the separation of the Weyl nodes in the momentum space. Namely, the anomalous Hall conductivity is expected to be proportional to the separation ($2k_{\mathrm{W}}$) with a sign opposite to the induced magnetization, as shown in the right-side formula in Eq. 1. Consistently, measured {\RyxA} and {\syxAint} are proportional to $M$ with the opposite sign, as confirmed in Figs. 2 and 3(a). The anomalous Hall effect is similarly observed also for other samples with different carrier densities (samples A and C). As summarized in Fig. 3(b), the absolute value of {\syxAint} after saturation increases with lowering of the Fermi level towards the Dirac node. This increase can be interpreted as the approach to a semiquantized value expressed in Eq. 1, which is further supported by following model calculations of {\syxA}.

In {\CA}, there exist only the Cd 5$s$ and As 4$p$ bands over a wide energy range of 400 meV around the Dirac node \cite{theoryCA}. A tight binding model constructed based on the first-principles calculation reproduces the band structure of the three-dimensional Dirac semimetal, as confirmed in Fig. 4(a). Figures 4(b) and (c) compare the change in the Dirac dispersion when adding an exchange interaction in the tight binding model. The degeneracy of the Dirac nodes is lifted with the exchange splitting and then the Weyl nodes (W$^{+}$ and W$^{-}$) emerge along the $k_{\mathrm{[112]}}$ direction.

Figure 4(d) plots Fermi level dependence of {\syxA} calculated by applying Eq. 1 to the {\it ab}-{\it initio} tight binding model. Consistent with the observed dependence, its absolute value increases with the Fermi level approaching the Dirac node energy. Then it is saturated around the semiquantized value $-2({k_\mathrm{W}}/{\pi})({e^2}/{h}) = - 9$ $\Omega^{-1}\mathrm{cm}^{-1}$, where the factor of 2 comes from the number of the Weyl node pairs. The increase to the semiquantized value starts from when the Fermi level is well above the Lifshitz point energy, also consistent with a theoretical calculation in the minimal model \cite{AHEinmagneticWeyl1}. While the Lifshitz transition may affect transport properties of {\CA} \cite{sigmaxx_suggested}, {\syxA} substantially remains even above the Lifshitz point energy. In general systems, it is difficult to distinguish how much the Weyl nodes contribute to observed anomalous Hall conductivity, because {\syxA} is obtained by integrating the contribution of {\Oz} for all occupied bands as seen in Eq. 1. In the case of {\CA}, on the other hand, {\Oz} only around the Weyl nodes is important in the momentum space over a wide energy range of the Fermi level ($\sim 200$ meV) in Fig. 4(d). All the quantities such as exchange coupling, distance between the split Weyl nodes, anomalous Hall conductivity, and induced tiny spin moment are linearly proportional to each other due to the simple band structure, as confirmed in Fig. 4(e). As the Fermi level reduces with approaching the original Dirac node energy, the Fermi surface shrinks in size and comes to exist only around the Weyl nodes in the momentum space. The absolute value of {\syxA} accordingly increases, and then starts to decrease after the Fermi level passes through the Dirac node energy. Therefore, the observed anomalous Hall effect is well captured by the theoretical calculations based on the simple electronic structure of {\CA}.

Regarding the difference of anomalous Hall conductivity values between the experiments and calculations, contribution of the Cd3As2 surface bands is worth considering, since the surface state thickness may be comparable to the proximitized one in the present heterointerface system. Quantitative evaluation of the surface contribution to the intrinsic anomalous Hall conductivity will be important in future study. Another possible origin is simply that the exchange interaction may be larger than the present assumption of $J_{\mathrm{ex}}=10$ meV in the calculations (Figs. 4(c) and 4(d)). Actually, all the quantities in Fig. 4(e) including {\syxA} almost linearly increase with {\Jex} up to as large as 50 meV (see Supplemental Material \cite{supplemental}). As the other possibility, longer proximitized thickness than 2 nm may give smaller {\syxAint} in the experimental analysis. However, this may not be so likely, because a typical value of the interface thickness is only about 2 nm in nonmagnetic topological insulator heterointerfaces \cite{EuSBS3}. Whatever the case, the present calculations simply identify the intrinsic origin of the observed anomalous Hall effect.

In summary, the above-room-temperature ferromagnetic state attributed to splitting of the Dirac nodes has been realized in the proximitized topological Dirac semimetal films. This has been first achieved by combining the simple Dirac dispersion and strong perpendicular magnetization in the heterostructure. Comparison to the theoretical calculations has also confirmed that the Fermi-level dependent anomalous Hall conductivity reflects the Berry curvature originating in the split Weyl nodes, as expected in the minimal model of the magnetic Weyl semimetals. The present Dirac-semimetal/ferromagnetic-insulator heterostructure can be an ideal platform for exploring unique spintronic properties and functions proposed for simple magnetic Weyl semimetals \cite{magneticWeylApp1, magneticWeylApp3}. Furthermore, topological semimetal heterostructures using such proximity effects will shed light on enigmatic transport phenomena observed in topological semimetals, including possible Weyl orbit motion \cite{weylorbit1, weylorbit2, weylorbit3, weylorbit4} and topological superconductivity \cite{SC1, SC2} appearing on the surface of {\CA}.

We acknowledge fruitful discussions with N. Nagaosa, Y. Tokura, S. Maekawa, Y. Araki, K. Yamamoto, E. Saitoh, S. Nakatsuji, J. Matsuno, Y. Yamaji, H. Ishizuka, and M. Hirschberger. This work was supported by JST PRESTO Grant No. JPMJPR18L2 and CREST Grant No. JPMJCR16F1, Japan, by Grant-in-Aids for Scientific Research (B) No. JP18H01866 from MEXT, Japan, and by TEPCO Memorial Foundation, Japan.

\clearpage

\begin{figure}
\begin{center}
\includegraphics*[width=16.0cm]{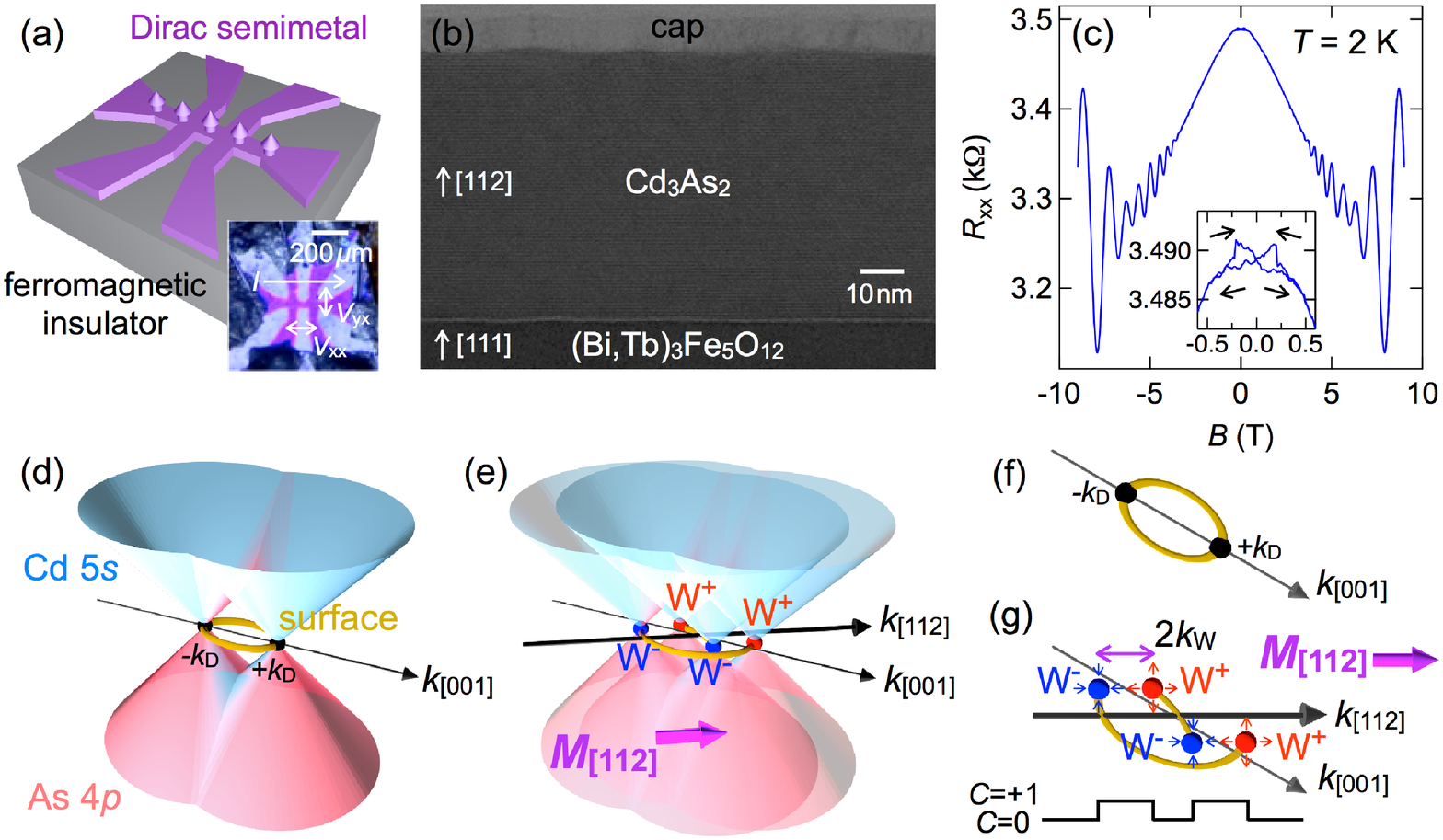}
\caption{Dirac-semimetal/ferromagnetic-insulator heterostructure. (a) Schematic illustration of a topological Dirac semimetal (purple) interfaced with a ferromagnetic insulator (gray). Magnetization is expected to be induced in the Dirac semimetal due to the magnetic proximity effect at the heterointerface. The inset shows an optical image of the heterostructure device. (b) Cross-sectional transmission electron microscopy image of the heterostructure, consisting of a 60 nm {\CA} film epitaxially grown on a Bi-substituted rare-earth iron garnet ({\BTIG}) substrate. The bright line seen between the film and substrate is due to an amorphous {\CA} layer formed by the irradiation of electron beams during the observation \cite{supplemental}. (c) Magnetoresistance with clear Shubnikov-de Haas oscillations, taken for the film (sample B) at 2 K. The inset is an enlarged view showing a butterfly shaped hysteresis loop around zero field. (d) Electronic structure of the topological Dirac semimetal {\CA}. Two Dirac nodes at $\pm k_{\mathrm{D}}$ are protected by crystal rotational symmetry along the [001] direction. (e) Evolution of the electronic structure caused by magnetization induced perpendicular to the (112) film plane. The Dirac nodes are split into paired Weyl nodes with opposite chirality (W$^{+}$ and W$^{-}$), at a distance of 2$k_{\mathrm{W}}$ along the $k_{\mathrm{[112]}}$ direction. The distance may be changed along the depth direction under a decaying exchange potential. (f) and (g) Corresponding top views at the Dirac node energy in (d) and (e). A Chern number profile along $k_{\mathrm{[112]}}$ is schematically shown at the bottom.
}
\end{center}
\end{figure}

\begin{figure}
\begin{center}
\includegraphics*[width=15.5cm]{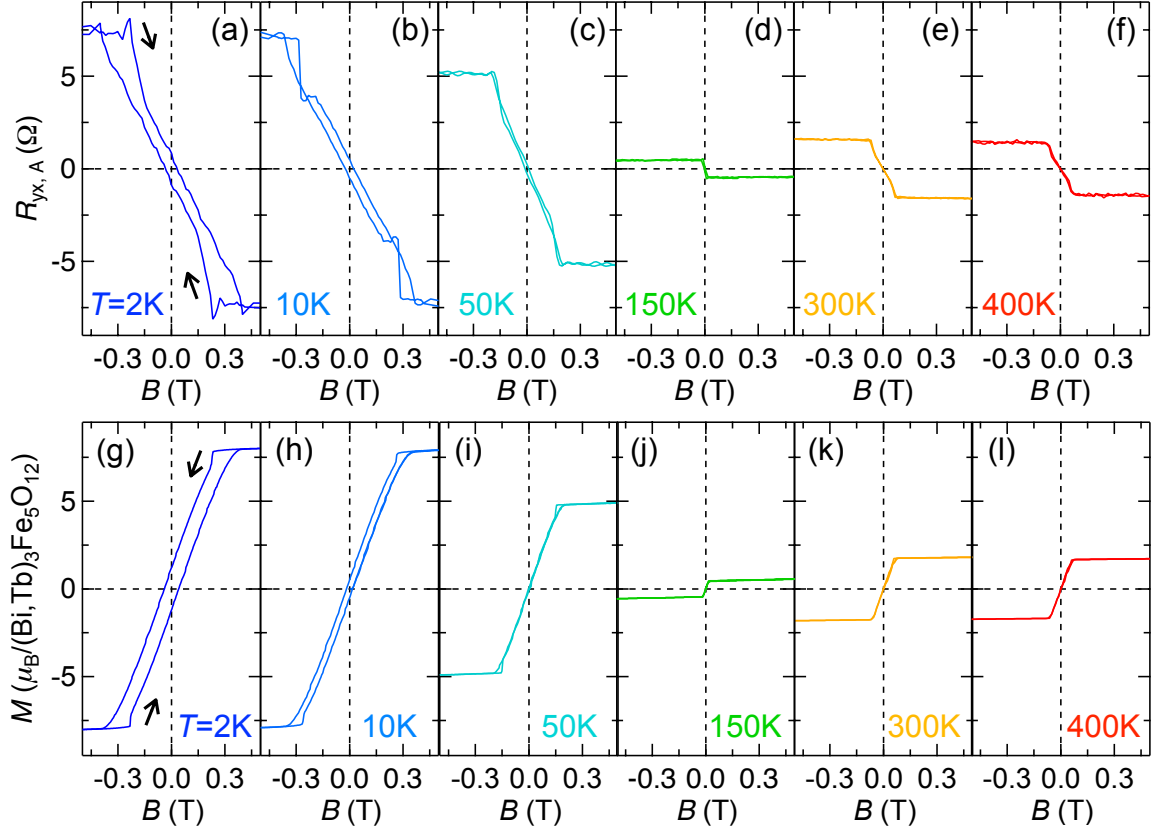}
\caption{Anomalous Hall effect observed in the heterostructure. (a)--(f) Anomalous Hall resistance and its temperature dependence taken for the {\CA}/{\BTIG} heterostructure (sample B) up to 400K. The arrows indicate the magnetic-field sweep directions. (g)--(l) Magnetization curves measured for the perpendicular direction of the garnet substrate at the same temperatures.
}
\end{center}
\end{figure}
\clearpage
\newpage

\begin{figure}
\begin{center}
\includegraphics*[width=13.0cm]{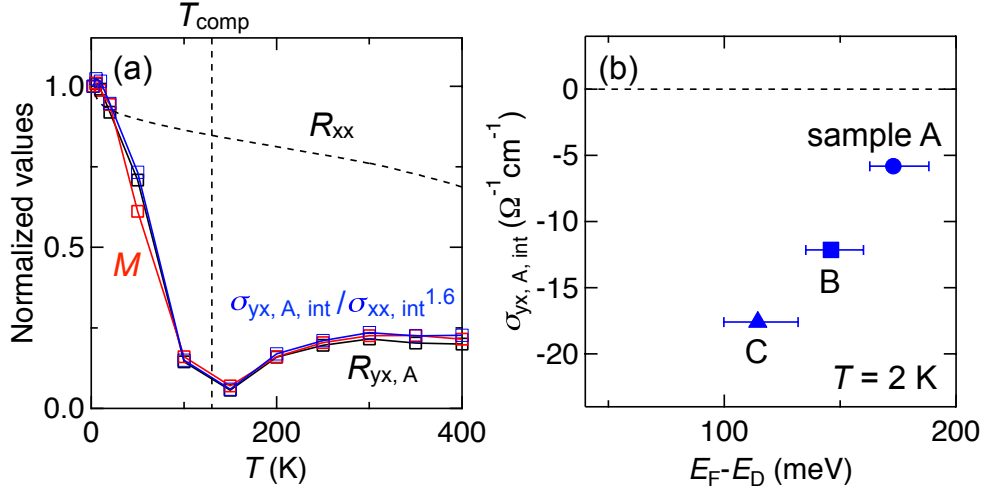}
\caption{Magnetization and Fermi-level dependence. (a) Longitudinal resistance and saturation values of anomalous Hall resistance, magnetization, and interface anomalous Hall conductivity are normalized at 2 K and plotted as a function of temperature (sample B). For comparison, {\syxAint} is also divided by {\sxxint}$^{1.6}$ for excluding an effect from the temperature change of the longitudinal conductivity in the dirty regime \cite{intrinsicAHE1}. The anomalous Hall and magnetization signals once vanish at the magnetic compensation temperature of the garnet substrate ($T_{\mathrm{comp}}=130$ K), while the longitudinal resistance exhibits monotonic semiconducting temperature dependence. (b) Fermi-level dependence of the interface anomalous Hall conductivity, observed as the saturation value for samples A, B, and C at 2 K. Here the interface thickness affected by the magnetic proximity is assumed to be 2 nm \cite{EuSBS3}. The Fermi energy {\EF} is estimated by analyzing the Shubnikov-de Haas oscillations and plotted from the Dirac node energy {\ED} \cite{PLDfilm1, supplemental}.
}
\end{center}
\end{figure}

\begin{figure}
\begin{center}
\includegraphics*[width=15.5cm]{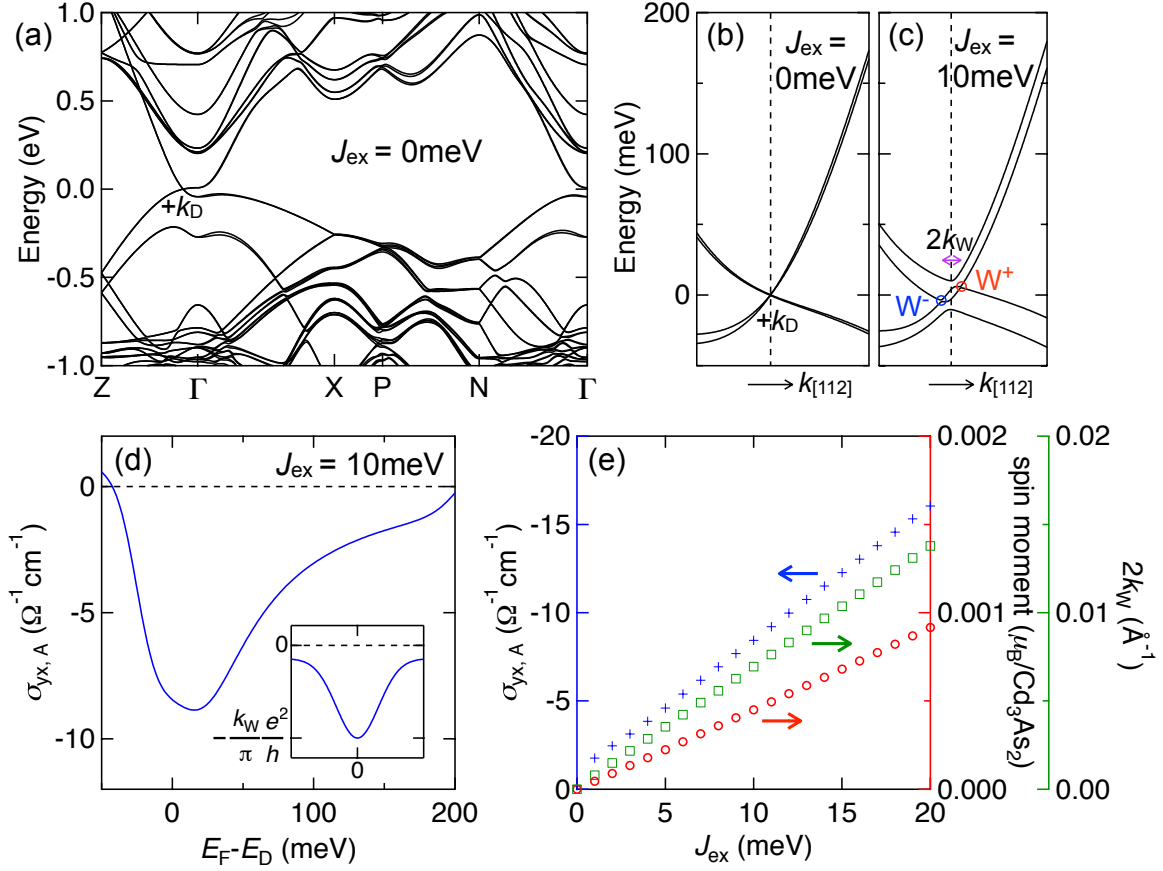}
\caption{Anomalous Hall conductivity calculated for ferromagnetic {\CA}. (a) Overall band structure of nonmagnetic {\CA}, reproduced by an {\it ab}-{\it initio} tight binding model. (b) and (c) Dirac dispersion plotted along the $k_{\mathrm{[112]}}$ direction and its change in a ferromagnetic state calculated with the exchange interaction of $J_{\mathrm{ex}}=10$ meV. (d) Fermi-level dependent {\syxA} calculated for the ferromagnetic state with the split Weyl nodes. The inset depicts the semiquantization of {\syxA} in the case of the minimal model \cite{AHEinmagneticWeyl1, AHEinmagneticWeyl2}. (e) {\Jex} dependence of {\syxA} at the Dirac node energy, induced spin moment per the formula unit, and separation of the Weyl nodes $2k_{\mathrm{W}}$.
}
\end{center}
\end{figure}

\end{document}